# A region segmentation based algorithm for building crystal position lookup table in scintillation detector

WANG Hai-Peng (王海鹏)[1,2,3]　　YUN Ming-Kai (贠明凯)[1,2]　　LIU Shuang-Quan (刘双全)[1,2]
FAN Xin (樊馨)[1,2,3]　　CAO Xue-Xiang (曹学香)[1,2]　　CHAI Pei (柴培)[1,2]　　SHAN Bao-Ci (单保慈)[1,2;1]

[1] Key Laboratory of Nuclear Radiation and Nuclear Energy Technology, Institute of High Energy Physics, Chinese Academy of Sciences，Beijing 100049, China

[2] Beijing Engineering Research Center of Radiographic Techniques and Equipment, Beijing 100049, China

[3] University of Chinese Academy of Sciences, Beijing 100049, China

**Abstract:** In scintillation detector, scintillation crystals are typically made into 2-dimension modular array. The location of incident gamma-ray need be calibrated due to spatial response nonlinearity. Generally, position histograms, the characteristic flood response of scintillation detectors, are used for position calibration. In this paper, a position calibration method based on crystal position lookup table which maps the inaccurate location calculated by Anger logic to the exact hitting crystal position has been proposed, Firstly, position histogram is segmented into disconnected regions. Then crystal marking points are labeled by finding the centroids of regions. Finally, crystal boundaries are determined and crystal position lookup table is generated. The scheme is evaluated by the whole-body PET scanner and breast dedicated SPECT detector developed by Institute of High Energy Physics, Chinese Academy of Sciences. The results demonstrate that the algorithm is accurate, efficient, robust and general purpose.

## 1 Introduction

Scintillation detectors have been widely applied in nuclear medical imaging systems such as positron emission tomography (PET) scanner, single photon emission computed tomography (SPECT) scanner to detect the position, energy and time information of gamma-ray. A scintillation detector module consists of a 2-dimension array of crystals coupled with photomultiplier tubes (PMTs) or position sensitive photomultiplier tubes (PSPMTs)[1]. These crystals are arranged in matrix and separated by opaque material to prevent scintillation photons from traversing between crystals. When a scintillation crystal undergoes gamma-ray interaction, scintillation photons are generated. Then the photons are collected and converted to electronic signals by the PMTs or PSPMTs[2]. Using the signals and Anger positioning algorithm[3], position coordinate (x, y) can be calculated to be as the gamma-ray incidence position. However, because of spatial response nonlinearity in the detection system, the calculated position coordinate cannot reflect the gamma-ray incidence position directly. The locating inaccuracy will impair the spatial resolution of the detector seriously[4]. So position calibration need to be done for correcting the mistaken location. Generally, a crystal position lookup table, mapping the calculated position coordinates to incident crystal indexes, is used for position calibration[5]. In order to build crystal position lookup table, a flood image which is the histogram of detected gamma-ray counts in each calculated position coordinate, is employed. An array of spots are shown in position histogram.

* Supported by National Natural Science Foundation of China (81101175), XIE Jia-Lin Foundation of Institute of High Energy Physics (Y3546360U2)
1) E-mail: shanbc@ihep.ac.cn

Each spot corresponds to a single crystal in the crystals array. So a crystal position lookup table can be built by segmenting the spot regions and labelling each region with corresponding crystal index.

At present, several approaches have been proposed to build crystal position lookup table using position histogram. A maximum likelihood estimation method based on Gaussian mixture models (GMMs) was presented by Stonger et al[6]. The approach decreases the probability of mistaken crystal identification, and is suitable for any possible crystal configuration. However, that is impractical for the large dimension position histogram, because a large number of parameters for the GMMs must be estimated. An approach to determining valley points as crystal boundary points in position histogram was proposed by Chai et al[7], which was also based on GMMs. This method is efficient in the case of small crystal array with uniform characteristic, but it is time-consuming to correct the crystal marking points interactively when the crystal number is large or the response of crystals varies dramatically. An approach based on neural networks and self-organizing feature maps was implemented by Lazzerini et al[8]. A similar method was proved effective on the Siemens Inveon detector by Hu et al[9]. The method can achieve high accuracy for their specific detectors, but suffer from instability due to inconsistencies in training data. A crystal identification method using non-rigid registration to a Fourier-based template was proposed by Chaudhari et al[10]. The method has a good effect on crystal identification, but it is time-consuming to generate a template image non-automatically. What's more, edge searching based algorithm and principal component analysis based algorithm were also proposed to build crystal position lookup table [11] [12].

In this work, a new general method is presented for building crystal position lookup table based on region segmentation algorithm. This approach is especially pertinent for the position histogram having large dimension and non-uniform intensity distribution, and has a high robustness in the position histogram with irregular spots position distribution.

## 2 Methods and materials

In this algorithm, each crystal marking point corresponding to a single crystal must be identified in position histogram first. And then crystal boundary points are determined by finding the centers of every four local neighboring crystal marking points. The position histogram is segmented into regions by connecting the adjacent crystal boundary points[7]. Each region is labeled by corresponding crystal index, which is the crystal position lookup table. The processing flow is shown as in Fig.1.

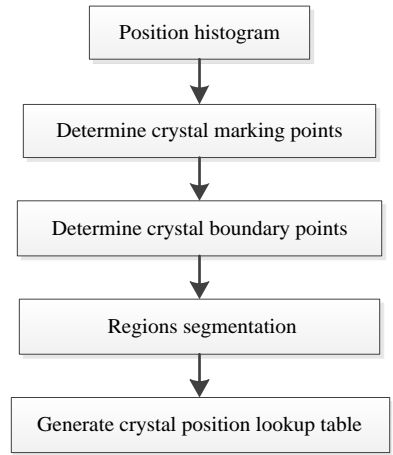

Fig.1. Processing flow for generating crystal position lookup table using position histogram

In the whole process, the key procedure is to identify crystal marking points accurately. Several image processing methods are used to reduce noise and enhance the contrast in position histogram image. And then the position histogram is divided into many disconnected regions by threshold segmentation. After that the centroid of each region is chosen as the corresponding crystal marking point. The processing steps are depicted in detail as follows.

**2.1 Noise reduction**

The position histogram acquired from detector includes much noise, which may give rise to mistaken segmentation. A mean filter with a 3×3 kernel is applied to minimize the noise. A larger kernel results in greater noise reduction, but also causes serious degradation of normal spots. In addition, clear background is performed in order to reduce the noise and accelerate the following procedures. The pixels that are less than a certain value are removed, and only the remaining pixels are processed by the following steps.

**2.2 Normalization**

Due to the difference of detection efficiency in detector module, the distribution of spots intensity is non-uniform in position histogram. It's hard to choose a uniform threshold for image segmentation. Normalization procedure is to make each local region have the similar intensity. After clearing background, the position histogram has been divided into several disconnected regions. And the average intensity of each region is denoted by $Al_i$, and $i$ is the index of region. The global average intensity of all regions is $Ag$. The normalization factor of $Ag / Al_i$ is multiplied by each pixel of the $i$-th region, which can adjust the position histogram to a uniform level of intensity.

**2.3 Histogram equalization**

Histogram equalization can increase contrast of position histogram by spreading out the frequent intensity values[13], and make spots have the similar size and intensity. This process is beneficial to the following threshold segmentation procedure. First, position histogram is scaled to a certain range, $L$, linearly. Then let $N_k$ be the number of pixels with intensity $k$, and $k$ has a range from 0 to $L - 1$. The total pixels are denoted as

$$N = \sum_{k=0}^{L-1} N_k \qquad (1)$$

Then the new pixel intensities are obtained by transforming the pixel intensities, $k$, by the function

$$T(k) = floor\left(L \sum_{i=0}^{k} \frac{N_k}{N}\right) \qquad (2)$$

Where $floor(\ )$ rounds down to the nearest integer.

**2.4 Threshold segmentation**

Through a series of processing, the position histogram has a better distribution, and all the spots have the similar shape and characteristic. The next procedure is to find an appropriate threshold to segment the position histogram. The principle of selecting the threshold is that the number of isolated regions segmented is as close as possible to the crystals number.

**2.5 Find crystal marking points**

The centroid of each segmented region is picked as the crystal marking point. The average of all pixels coordinate of the region is calculated to determine the position of central pixel. However, it is not sure that all the crystal marking points are identified accurately. It is more likely to miss some regions using higher segmentation threshold, or merge some regions when a relative lower threshold is selected. So interactive correction is carried out, which will make sure that all the crystal marking points are identified accurately.

**3 Result**

The algorithm was implemented in C programming language, and integrated in the quality control module of PETCTview software system developed by Institute of High Energy Physics, Chinese Academy of Sciences. The graphical user interface (GUI), shown in Fig.2, was designed and implemented in Qt, which can guide us to complete all the procedures conveniently. The GUI also provides the function of acquiring position histogram, displaying image, correcting crystal marking points interactively, and uploading the crystal position lookup table to the device. What's more, the algorithm was verified in the breast dedicated single photon emission tomography system (SPEMi) which was also developed by Institute of High Energy Physics, Chinese Academy of Sciences.

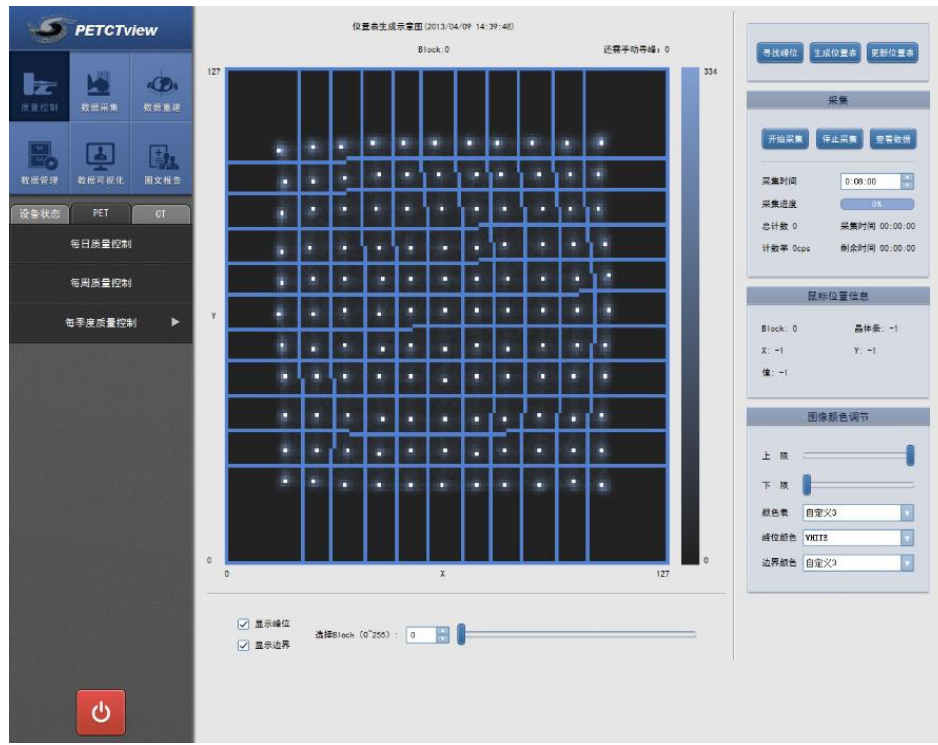

Fig.2. GUI for building crystal position lookup table implemented in the quality control module of PETCTview software system

**3.1 Building crystal position lookup tables in whole-body PET detector**

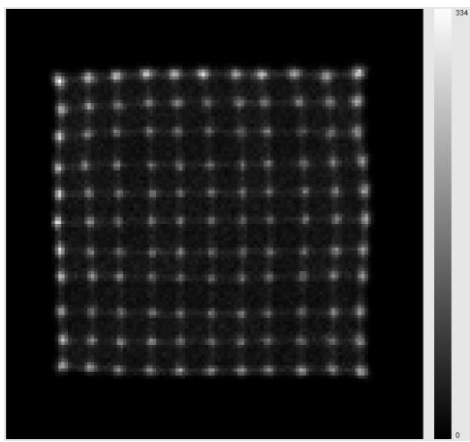

Fig.3. A position histogram image using the Ge-68 rod source acquired in whole-body PET detector

For the whole-body PET scanner, there are 256 detector blocks in total, and each block consists of an 11×11 array of LYSO crystals coupled with H8500 PMT. Each crystal element has the size of 3.5mm×3.5mm×25mm. The position histograms are obtained using Ge-68 rod source revolving around the scanner ring, and has a resolution of 128×128 pixels. The position histogram of detector block 0 is shown in Fig.3, which indicates that 11×11 spots are distinguished clearly.

The processing and segmentation results of the position histogram are presented in Fig.4 (a) ~ (d). Brighter pixels represent higher intensity levels. The result of finding crystal marking points is shown in Fig.4 (e), and green points represent crystal marking points. The result of building crystal boundaries is shown in Fig.4 (f), and blue lines represent crystal boundaries.

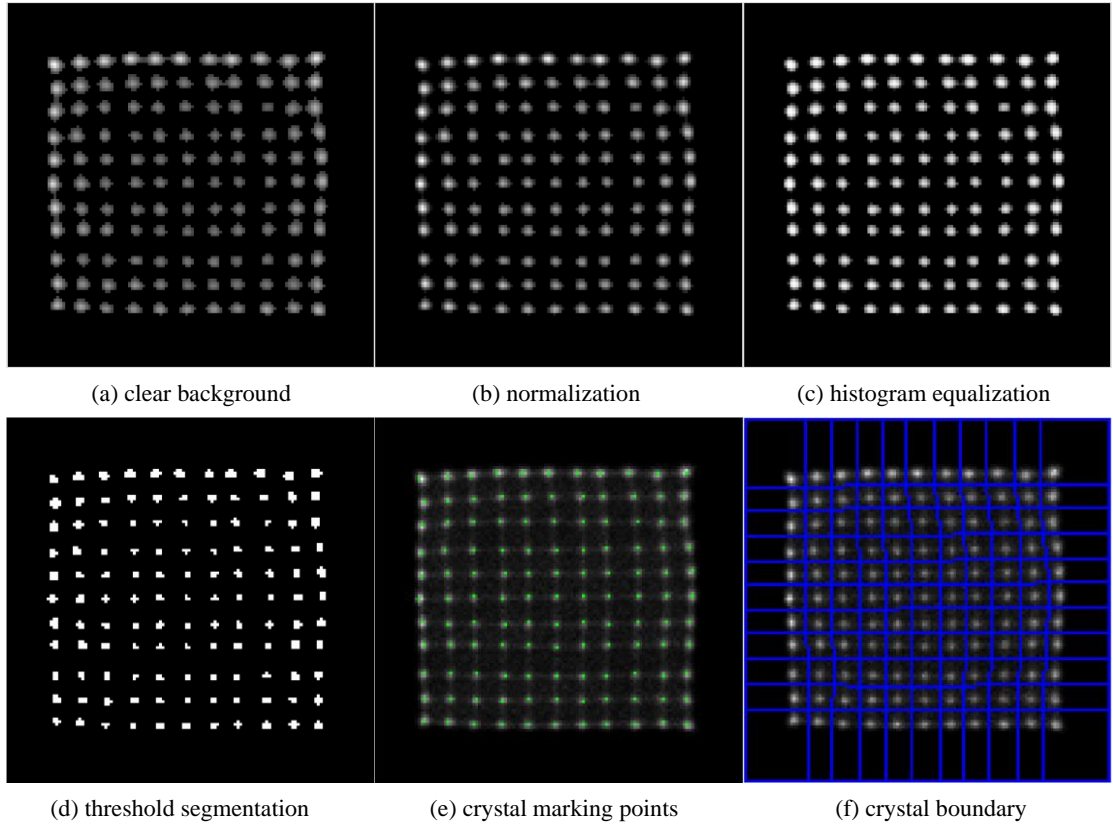

(a) clear background  (b) normalization  (c) histogram equalization

(d) threshold segmentation  (e) crystal marking points  (f) crystal boundary

Fig.4. Processing results using region segmentation based algorithm in whole-body PET detector

## 3.2 Building crystal position lookup table in SPEMi detector

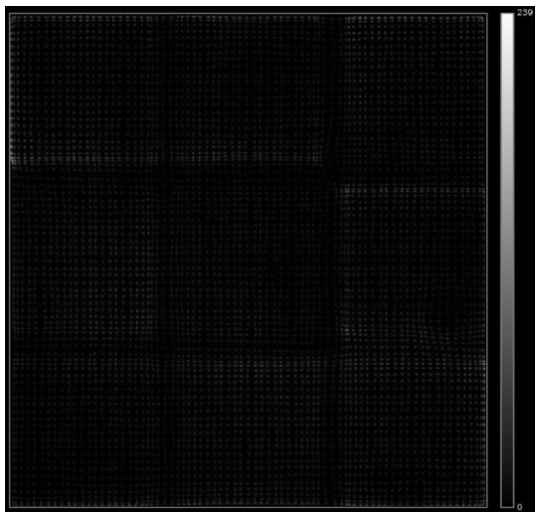

Fig.5. A raw position histogram image of the Tc-99m flood source acquired in SPEMi detector

The SPEMi detector module consists of a NaI crystals array and nine H8500 PMTs array. The crystals array is of a 77×77 matrix with the element size 1.8mm×1.8mm×6mm. The position histogram measured by SPEMi detector using Tc-99m flood source is shown in Fig.5, which has a resolution of 1024×1024 pixels. The result indicates that only 75×75 spots are distinguished, which is due to the adjacent rows and columns merging together in the borders of position histogram.

The processing and segmentation results of the position histogram are presented in Fig.6 (a) ~ (d). The result of finding crystal marking points is shown in Fig.6 (e). The result of building crystal boundaries is shown in Fig.6 (f).

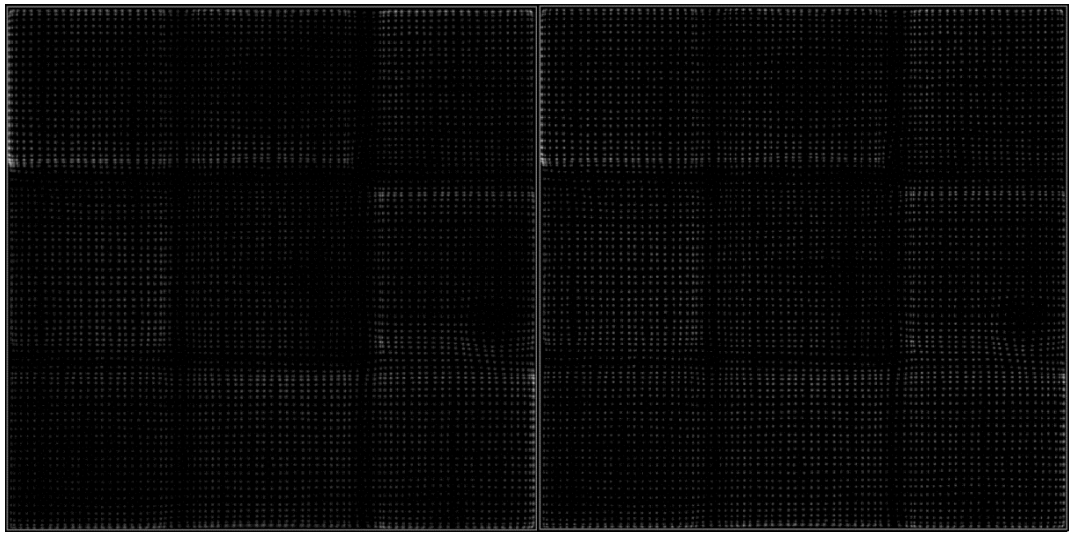

(a) clear background  (b) normalization

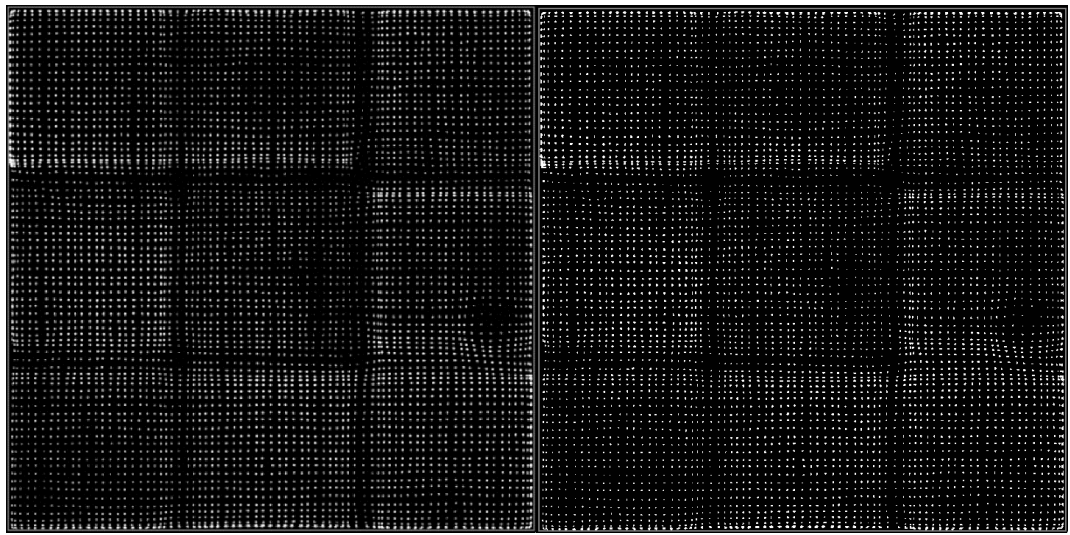

(c) histogram equalization  (d) threshold segmentation

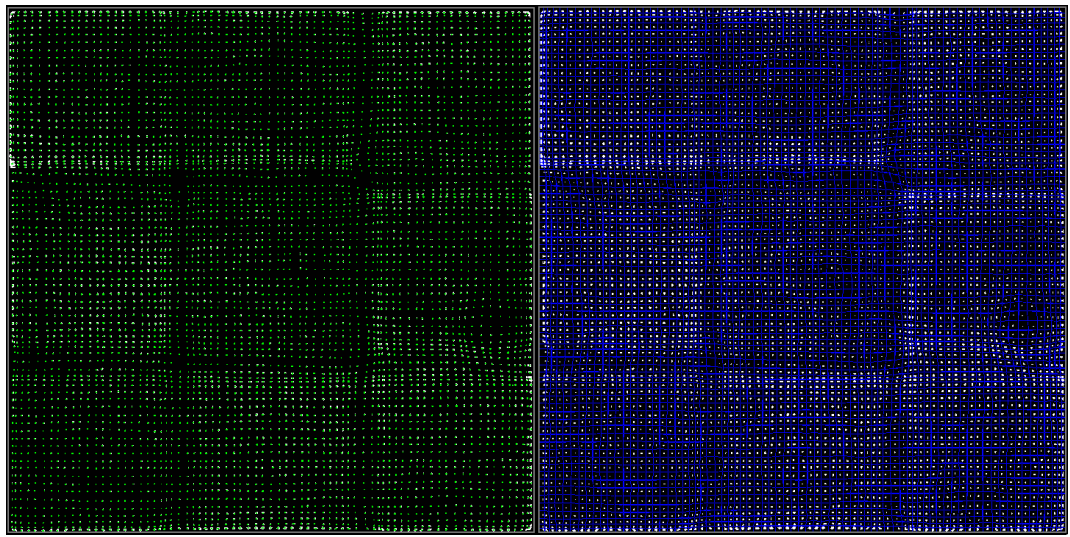

(e) crystal marking points  (f) crystal boundary

Fig.6. Processing results using region segmentation based algorithm in SPEMi detector

## 4 Discussion

The approach of building crystal position lookup table can broadly be divided into three processes. (i) Segment the position histogram into disconnected regions. (ii) Determine the crystal marking points. (iii) Determine the crystal boundaries. The segmentation process is the most critical, which will directly affect the efficiency of building crystal position lookup table. The accurate crystals identification will decrease the consuming time of interactive correction. So the percentage of identification of crystal marking points is as the evaluation of the algorithm.

To the position histograms of whole-body PET detectors, all the crystal marking points in 256 detector blocks have been identified accurately, and the interactive correction procedure can be skipped. It takes no more than 2 seconds to build the $11\times11\times256$ crystals position lookup table in Intel(R) Core(TM) i5-2400 3.10GHz CPU.

Among the distinguishable 75×75 spots of SPEMi position histogram, 99.23% crystal marking points are identified accurately, and only 59 crystal marking points, including 16 mistaken location points and 43 missing location points, need to be corrected interactively by clicking the mouse. The mistaken and missing marking points appear primarily in the borders due to the merging of two adjacent spots, and in the joint parts of PMTs due to the dead gaps. These regions of position histogram have bad distribution characteristic.

For different characteristic position histograms, it is necessary to adjust the segmentation threshold subtly to get better segmentation effect. That will shorten the interactive correction time and increase efficiency. In this work, many trials are performed with the threshold changing in descending order among a certain range, and the number of segmentation regions goes up first, and then declines with the decreasing of the threshold. The first threshold of closest total crystals number is picked as the final segmentation threshold.

In the position histogram of whole-body PET detectors, the intensity levels are scaled from 0 to 255. The selection of scaling range depends on the quality of position histogram. In general, a large scaling range needs to be chosen when the position histogram has a bad distribution characteristic. The regions number of segmentation changing with different thresholds in detector block 0 is shown in Fig.7. The 199 is chosen as final segmentation threshold. The results of segmentation with different thresholds are shown in Fig.8. As to SPEMi detector, the intensity levels are scaled from 0 to 1023. The similar results are shown in Fig.9 and Fig.10, and the 537 is chosen as final segmentation threshold.

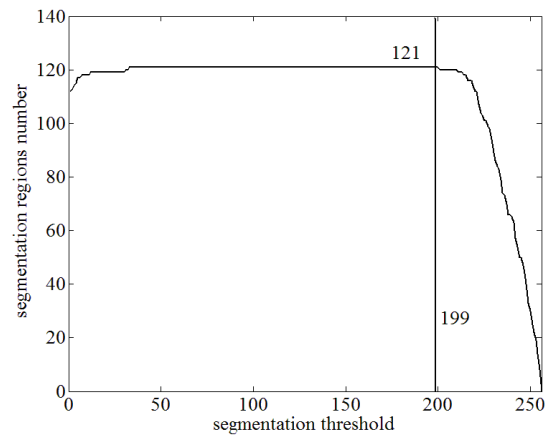

Fig.7. Regions number of segmentation changing with different thresholds in detector block 0 of PET

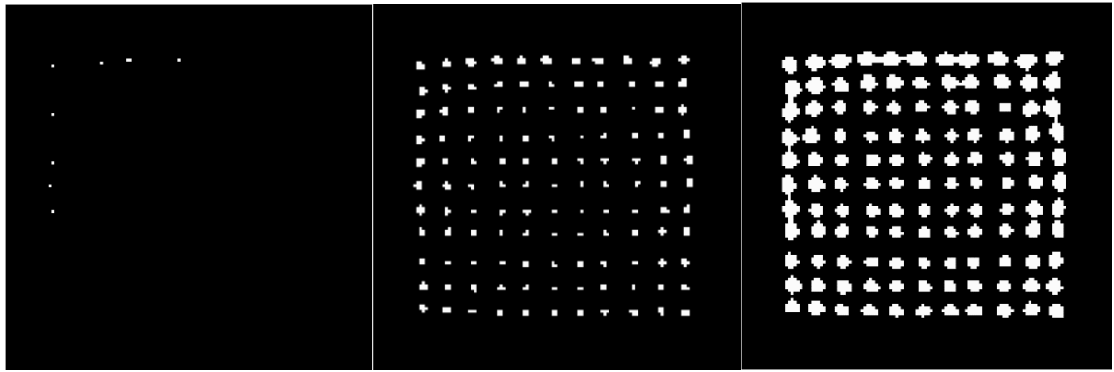

(a) threshold = 255　　　　(b) threshold = 199　　　　(c) threshold = 0

Fig.8. Segmentation results with different thresholds in detector block 0 of PET

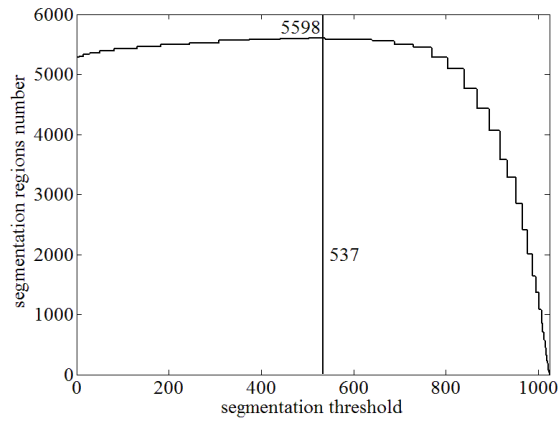

Fig.9. Regions number of segmentation changing with different thresholds in SPEMi detector

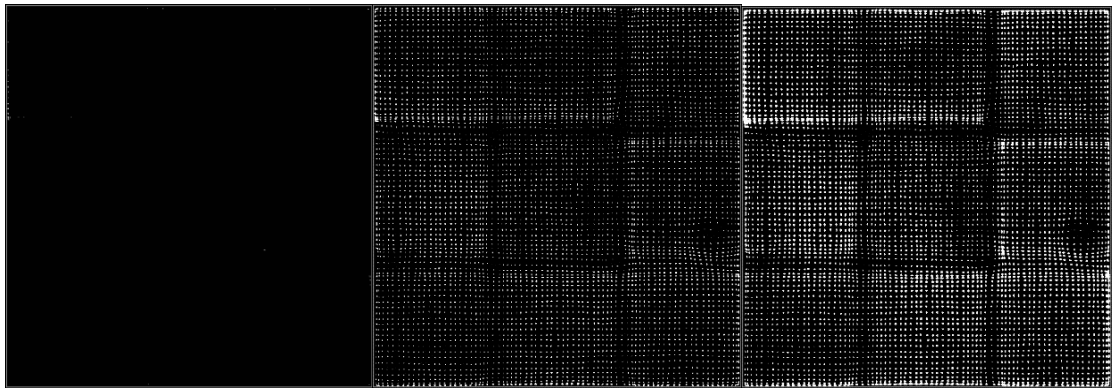

(a) threshold = 1023　　　　(b) threshold = 537　　　　(c) threshold = 0

Fig.10. Segmentation results with different thresholds in SPEMi detector

## 5 Conclusion

In this paper, a region segmentation method has been presented for building crystal position lookup table in scintillation detector. Since the method does not rely on specific characteristics of detectors, it can be adapted to work with all kinds of scintillation scanners, and not restricted to the dimension of crystals array. It was proved to be highly accurate, efficient and robust in the whole-body PET detector and SPEMi detector developed by Institute of High Energy Physics, Chinese Academy of Sciences. Moreover, the method is effective in a variety of position histograms, and has an advantage in the position histogram with large dimension, non-uniform intensity distribution or a high degree of distortion.